\documentclass[aps,pra,twocolumn,showpacs]{revtex4}
\usepackage{times,amsfonts}
\usepackage{graphicx}
\begin{document}
\title{Are the Tonks  regimes 
 in the continuum and on the lattice truly equivalent?}
\author{M.~A. Cazalilla}
\affiliation{Donostia International Physics Center (DIPC), Paseo Manuel de Lardizabal, 4, 
20018-Donostia (Spain).}
\begin{abstract}
  Motivated by recent experiments, we compare the Tonks (\emph{i.e.} hard-core boson gas) regime
 achieved in an optical lattice with the Tonks regime of a one-dimensional Bose gas in the continuum. For the lattice gas,   
we compute  the local (\emph{i.e.} on-site) two-body correlations as a function
of  temperature  and the filling of the lattice. It is found that this function saturates to a constant value with increasing temperature.  Furthermore, the parameter that characterizes the   long-distance  correlations in the lattice Tonks regime is also obtained, showing that on the lattice the long-distance correlations enter the 
Tonks regime more rapidly than in the continuum.
\end{abstract}
\pacs{05.30.Jp, 3.75.Hh, 3.75.Lm}
\maketitle

  Strongly interacting 
 gases of bosons in one dimension have been recently realized 
 using optical lattices~\cite{S04,P04,L04}. By loading a Bose-Einstein
 condensate into a deep two-dimensional optical lattice, an array of one-dimensional
 atomic systems  (tubes) was created~\cite{L04,2DOL}. Strong correlations amongst the bosons were subsequently
 induced by turning on a third  lattice along the axis of the tubes, which further decreases the ratio
 of  kinetic  to interaction energy~\cite{S04,P04}. These anisotropic optical lattices exhibit a rich
 phase diagram~\cite{S04,H04}. In particular, by making the third lattice deeper and reducing the  filling 
 of the lattice below one particle per site, the Tonks regime, where the bosons effectively become hard-core
 and behave in many respects like fermions, was   reached in the experiments
 reported in  Ref.~\onlinecite{P04}. Time of flight measurements
 of the momentum distribution  showed good agreement with a fermionization approach 
 which accounted for finite temperature, finite-size, and trap effects~\cite{P04}.
\begin{figure}[t]
\vskip -12mm
\begin{center}
\includegraphics[width=\columnwidth]{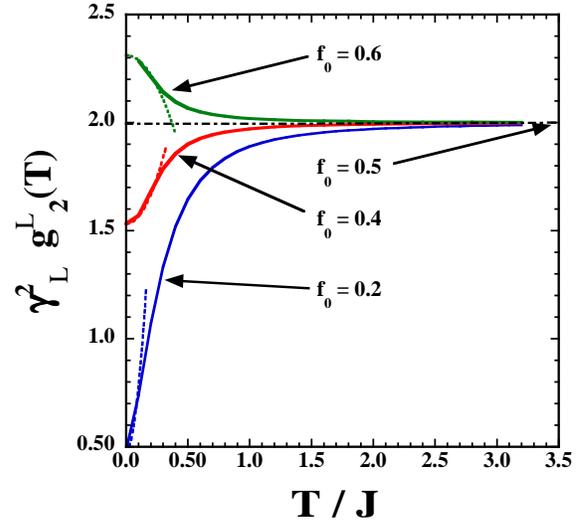}
\caption{Local two-body correlation function $g^{L}_2(T, f_0) = f^{-2}_0 \langle (b^{\dag}_m)^2 (b_m)^2\rangle $ 
(times the square of $\gamma_L \equiv U/J \gg 1$ )  in the lattice Tonks regime, 
as a function of temperature $T$ in units of the hopping $J$,  for  several lattice fillings $f_0 < 1$. 
At temperatures $T \gg J$, $g^L_2(T,f_0)$ saturates to the value at  half-filling, 
which is temperature independent. This behavior should be contrasted with the monotonic increase
at high  temperatures of  the corresponding function for the continuum  (\emph{i.e.} Lieb-Liniger) model: 
$g_2(T) =  \rho^{-2}_0 \langle (\Psi^{\dag}(x))^2 ( \Psi(x))^2 \rangle \propto T$ (see Refs.~\onlinecite{K02,C03}).
The dashed lines correspond to the low-$T$ analytical approximation, Eq.~(\ref{sommerfeld}).\label{fig1}}
\end{center}
\end{figure}

 	The achievement of the Tonks regime in an optical lattice  raises 
a number of questions about the equivalence  of this system with the continuum Tonks regime.  
Mathematically speaking,
the two types of Tonks gases correspond to the strongly interacting limit of two physically different models.
The lattice Tonks gas (LTG) is obtained from the Bose-Hubbard model ($m=1,\ldots,M_0$),
\begin{equation}
H_{\rm BH} = -\frac{J}{2} \sum_{m} \left( b^{\dag}_{m+1} b_m + {\rm H.c.} \right) + \frac{U}{2}  \sum_{m}  (b^{\dag}_m)^2 (b_m)^2\label{BH}
\end{equation}
in the regime where $\gamma_L \equiv U/J \gg 1$~\footnote{Our definition of $\gamma_L$ corresponds to the $\gamma$ 
used in Ref.~\onlinecite{P04}. Note  that our conventions for $U$ and $J$ 
differ by  factors of $1/2$  from the convention used in~\cite{P04}, but they cancel
out after taking the ratio.} and 
the filling of the lattice $f_0 = N_0/M_0 < 1$ ($N_0$ being the number
of atoms and $M_0$ the number of lattice sites). However, the continuum 
Tonks gas (CTG) regime is obtained from the Lieb-Liniger model~\cite{LL63},
\begin{equation}
H_{\rm LL} =  \int^{L}_0 dx  \, \frac{\hbar^2}{2M}  \big|\partial_x \Psi (x) \big|^2 + \frac{g}{2} (\Psi^{\dag}(x))^2 (\Psi(x))^2 
\label{LL}
\end{equation}
when the  parameter $\gamma \equiv  Mg/\hbar^2 \rho_0 \gg 1$~\cite{LL63,D01}. The presence of the 
density $\rho_0 = N_0/L$ in this parameter  is a distinct feature of the continuum model. It means that the CTG regime can be
reached either by increasing the interaction coupling $g$ or by decreasing the density, $\rho_0$. By contrast,
the LTG regime can be reached  by increasing the ratio $\gamma_L = U/J$  at any value of the filling $f_0$.
For low temperatures and  fillings (\emph{i.e.}  $f_0 \ll 1$) the two
Tonks regimes coincide since the Lieb-Liniger model   emerges as a low-density limit of the Bose-Hubbard
model (see \emph{e.g.} Ref.~\cite{C03}).
 However, in ~\cite{C03} it was  shown  that, provided one has \emph{at most} one particle per site (\emph{i.e.}
$n_{m} \leq 1$) and $U\gg J$, the 
Bose-Hubbard model can be effectively replaced by the following interacting fermion model:
\begin{eqnarray}
H_{\rm F} &=& -\frac{J}{2} \sum_{m} \left( c^{\dag}_{m+1} c_m + {\rm H.c.} \right) + H_1 + H_2, \label{hferm} \\
H_1 &=&   \,\, \frac{J}{2} \lambda_1  \sum_{m} \left( c^{\dag}_{m+1} n_{m} c_{m-1} + {\rm H.c.}\right), \label{corrhop}\\
H_2 &=&   -   J \lambda_2  \sum_{m} n_{m} n_{m+1}. \label{xxz}
\end{eqnarray}
The couplings $\lambda_1 = \lambda_2 = \gamma^{-1}_L = J/U $ are small, and the fermions are `almost' non-interacting. 
In this sense, therefore, one can speak of `fermionization' of  bosons. In this paper, we consider the 
regime where $\gamma$ or $\gamma_L$ are large, which should be of interest for current~\cite{S04,P04,L04} 
and future experiments exploring these correlated systems.

	The physical differences between the Tonks regimes of the Lieb-Liniger and Bose-Hubbard model can
only be addressed by an explicit calculation of their correlation properties. These fall into two classes: short distance
correlations are \emph{non-universal} (\emph{i.e.} model dependent), and therefore are expected to  be different 
in the LTG and the CTG regimes. On the other hand, long-distance (or small momentum) correlations are characterized 
by the same power-laws \emph{exactly at} the Tonks
limit: $\gamma$  or $\gamma_L \to +\infty$. Thus, 
for instance, in the thermodynamic limit at zero temperature, the momentum distribution
$n(p) \sim p^{-1/2}$ for $p \ll \rho_0$~\cite{L64,C02}. 
Nevertheless,  for finite values of $\gamma$ or $\gamma_L$,  
long-distance 
correlations are characterized by the  Luttinger-liquid parameter $K$ (\emph{e.g.}
$n(p) \sim p^{1/2K-1}$),  which is non-universal~\cite{C04b}.
Precisely, these non-universal features in the two Tonks regimes are what interests us
here. In this regard, it is important to notice that  for the Bose-Hubbard model 
there is no exact (\emph{i.e.} Bethe-ansatz) solution available~\cite{Ch82}, and therefore
\emph{analytical} results for non-universal properties are scarce.
In what follows, we  have summarized our results:

$\bullet$ In the LTG regime we have obtained the temperature and filling-fraction dependence of the on-site 
two-body correlation function $g^L_2(T,f_0) = f^{-2}_0 \: \langle (b^{\dag}_m)^2 (b_m)^2 \rangle$.
This is the lattice counterpart of the continuum-model $g_2(T) = \rho^{-2}_0 \langle (\Psi^{\dag}(x))^2 ( \Psi(x))^2 \rangle$. Both functions   vanish for $\gamma,\gamma_L \to +\infty$, that is, when fermionization is complete.  
Whereas  $g_2(T)$ was computed in Ref.~\cite{K02} for all $\gamma$ values, to the best of 
our knowledge no results existed for $g^L_2(T, f_0)$.  For  $\gamma_L \gg 1$ and  temperatures $T \ll U$,
we find:
\begin{equation}
g^L_2(T,f_0) =  2 \gamma^{-2}_L \left( 1 - \frac{f_2(T)}{f_0} \right) + O(\gamma^{-3}_L), \label{g2}
\end{equation}
where $f_2(T) = \langle c^{\dag}_{m+2} c_{m} \rangle$ (see below for details).  At half-filling, 
the non-interacting fermion system is invariant under particle-hole symmetry: $c_m \to (-1)^m c^{\dag}_m$,
which implies that $f_2(T) = 0$ at all temperatures. Thus we obtain the result that
$g^L_2(T,f_0=1/2) =   2 \gamma^{-2}_L$, independent of $T$. Results for arbitrary temperatures 
and several fillings are shown in fig.~\ref{fig1}. Although $g^{L}_2(T,f_0)$ is not directly related 
to the photo-association (PA) rate
in the lattice because of the overlap between the Wannier orbitals at different sites, 
it should be experimentally accessible by suddenly ramping up the  optical lattice  
before PA is performed in a time scale shorter than the atom tunneling time. After  a substantial increase
of   lattice depth, overlap between  the Wannier orbitals should become negligible.

$\bullet$ We have obtained the  Luttinger-liquid parameters $K$ and $v_s$  for fillings $f_0 < 1$ to leading
order in $\gamma^{-1}_L$:
\begin{eqnarray}
K &\simeq& 1 +  4 \gamma^{-1}_L \sin \pi f_0/\pi, \label{K} \\
v_s/v_F &\simeq& 1 - 4 \gamma^{-1}_L \: (f_0 \cos \pi f_0). \label{vs} 
\end{eqnarray}
where the Fermi velocity $v_F = J a \sin \pi f_0/\hbar$. 
Thus we conclude that  the Tonks regime is
more easily reached, as far as long distance correlations are concerned,
on the lattice than in the continuum. To see this, consider for instance a half-filled
lattice ($f_0 = 1/2$). Using the above formula for $\gamma_L = 10$, $K \simeq 1.13$, whereas
for the Lieb-Liniger model~\cite{C04b} 
$K \simeq 1 + 4/\gamma = 1.4$ for $\gamma = \gamma_L = 10$ (Indeed, 
$\gamma = 10$ seems  harder to achieve experimentally than $\gamma_L = 10$).
The fact that long-distance correlations rapidly become Tonks-like for relatively shallow 
lattices justifies the fermionization treatment used in Ref.~\cite{P04}.

		Next, we provide further details on the derivation of the above results. The key point is to notice
that for temperatures $T \ll U$ and filling less than one particle per site, the Bose-Hubbard model, Eq.~(\ref{BH}),
can be effectively replaced by the fermionic model  of Eq.~(\ref{hferm}), $H_{\rm F}$. In particular, the replacement
can be made for computing the (low-temperature) partition function  of (\ref{BH}):
\begin{equation}
Z = {\rm Tr} \, e^{-\beta ( H_{\rm BH} - \mu N)} = Z_0\: \Big\langle  {\cal T} e^{-
\int^{\hbar \beta}_{0} \frac{d\sigma }{\hbar} \: H_{\rm int}(\sigma)}\Big\rangle, \label{partition}
\end{equation}
where $\beta = 1/T$ and $H_{\rm int} = H_1 + H_2$ (cf. Eqs.~(\ref{corrhop}, \ref{xxz})). In the second 
expression $Z_0 = {\rm Tr}  \,  e^{-\beta(H_0 - \mu N)}$  and  
$\langle \ldots \rangle = {\rm Tr} \left[ \rho_{0}(\mu,\beta) \ldots \right]$, with
$\rho_0(\mu,T) = e^{-\beta(H_0 - \mu N)}/Z_0$,  and $H_0 = H_{\rm F} - H_{\rm int}$.
We can now expand $Z$ to the the lowest order in $\gamma^{-1}_L$ and 
obtain~\footnote{Note that it makes no sense to go beyond $O(\gamma^{-1}_L)$, as 
$H_{\rm F}$ is  the lowest order term in an expansion in powers of $J/U = \gamma^{-1}_L$
(see Ref.~\onlinecite{C03}).}, $\log (Z/Z_0) = - \beta \langle H_{\rm int} \rangle + O(\gamma^{-2}_L)$.
Thus we need to compute the thermal average of $H_{\rm int}$, which can be
readily done with the help of Wick's theorem. The result can be written as follows:
\begin{eqnarray}
\langle H_{\rm int} \rangle &=&  
\frac{1}{2} \lambda_1 J M_0 \left[ f_0\left(f_{+2} + f_{-2} \right) - \left( f^2_{+1} + f^{2}_{-1} \right)\right] \nonumber \\
&&- \lambda_2 J M_0 \left[ f^2_0 - f_{+1} f_{-1} \right] \label{hf}
\end{eqnarray}
where we have denoted ($l=0,\pm 1,\pm2$):
\begin{equation}
f_l(T) = \langle c^{\dag}_{m+l} c_{m} \rangle =  \frac{1}{M_0} \sum_{p} e^{-i  p l a}\: n(\epsilon_0(p),z),\label{fl}
\end{equation}
the function $n(\epsilon,z) = \left[z^{-1}e^{\beta \epsilon} +1 \right]^{-1}$ is the Fermi-Dirac
distribution for a fermion gas of  fugacity 
$z\equiv e^{\beta \mu}$;
$\epsilon_0(p) = -J \cos (pa)$  is the single-particle dispersion.  Interestingly, all the above results
follow from this simple expression, Eq.~(\ref{hf}).

 We begin by describing the calculation of $g^{L}_2(T, f_0)$. 
In an analogous manner to the continuum case~\cite{K02,C03},  
this function can be obtained using the Hellmann-Feynman theorem:
\begin{eqnarray}
g^L_2(T,f_0) &=& - \frac{2 f^{-2}_0}{M_0\beta} \: \frac{\partial}{\partial U} \log Z  \\
 &=&  
2 f^{-2}_0 \frac{\partial}{\partial U} \left(  \frac{\langle H_{\rm int} \rangle}{M_0}\right) + O(\gamma^{-3}_L).
\end{eqnarray}
By setting $\lambda_1 = \lambda_2 = \gamma^{-1}_L$ in (\ref{hf}), and assuming periodic boundary 
conditions~\footnote{This requires $N_0$ to be odd, $M_0$ to be even
and the gound state is non-degenerate.} so that $f_l = f_{-l}$, one  obtains
the first result given above, Eq.~(\ref{g2}). An alternative expression for $g^L_2(T,f_0)$
can be obtained after recasting
\begin{equation}
f_2(T) =  f_0 \left( 2J^{-2 }\overline{\epsilon^2}(T,z)  - 1 \right),
\end{equation}
where $ \overline{\epsilon^2}(T,z)  =  f^{-1}_0 \int d\epsilon\:  \epsilon^2 g(\epsilon) n(\epsilon,z)$, being
$g(\epsilon) = 1/\pi \sqrt{J^2 - \epsilon^2}$ the single-particle density of states. The advantage
of this form of $f_2(T)$ is that  the Sommerfeld expansion can be used
to extract the low-temperature behavior:
\begin{equation}
g^L_2(T,f_0) =g^{L}_l(T=0,f_0) +  \frac{4\pi}{3 \gamma^2_L}\frac{  (T/J)^2}{f_0 \tan \pi f_0} + O(\gamma^{-3}_L), 
\label{sommerfeld}
\end{equation}
and $g^L_2(T = 0,f_0) = 2\gamma^{-2}_L \left(1 - \sin(2\pi f_0)/2\pi f_0 \right)$. It is 
worth noticing that in the the low-filling limit $f_0 \to 0$ one recovers, from the above
expression, the asymptotic expression for the Lieb-Liniger gas obtained in Ref.~\onlinecite{K02}
(see also~\cite{C03}), provided one makes the following identifications between  the parameters
of both models, $M \to \hbar^2/Ja^2$ (\emph{i.e.} the effective mass), $g \to Ua$, $\rho_0 \to f_0/a$
(see~\cite{C03}). 

	Finally, let us discuss some interesting properties  of 
$f_0(T,z)$ and $f_2(T,z)$ defined in (\ref{fl}),  and their implications for $g_2(T,f_0)$. 
The first property is a consequence of the particle-hole symmetry of the
non-interacting spectrum, which implies that $f_0(T, z) + f_0(T, z^{-1}) = 1$, and 
hence that the fugacity for filling $1-f_0$ is the inverse of the fugacity for filling $f_0$.
Likewise, one can show that $f_2(T, z) + f_2(T, z^{-1}) = 0$, which implies that in practice
it suffices to compute $f_2(T,z)$ for fillings $f_0 \leq 1/2$.  
The other property of these
functions explains the saturation of $g^L_2(T,f_0)$  with increasing temperature
observed in fig.~\ref{fig1}: for $T\gg J$, it can be shown that  
$z \to f_0/(1-f_0)$ and $f_2(T,z) \to 0$; hence the local two-body correlation function
$g_2(T,f_0) \to g_2(T, f_0 = 1/2)$ at all $f_0 < 1$. In the end, this is 
a consequence of the finite number of degrees of freedom available on the
lattice.

  We finally consider the non-universal aspects of long-distance correlations, which
are parametrized by the Luttinger-liquid parameters $K$ and $v_s$. 
In order to obtain the latter, it is convenient to work with
the density and phase stiffness~\cite{C04b}:
\begin{eqnarray}
v_J = \frac{\pi M_0 a}{\hbar} \frac{\partial^2 E_0(N_0)}{\partial \phi^2}\Big|_{\phi = 0}
\quad v_N = \frac{M_0 a}{\pi \hbar} \frac{\partial^2 E_0(\phi=0)}{\partial N^2}\Big|_{N_0}\label{stiff}
\end{eqnarray}
The angle $\phi$ corresponds to a twist in the boundary conditions:
$c_{m+M_0} = e^{i \phi} c_m$.  This makes $f_{-l}(T=0) \neq f_{l}(T=0)$
for $l\neq 0 $, and an explicit evaluation at finite size and zero temperature 
yields:
\begin{equation}
f_l(T = 0,\phi) = \frac{e^{-i \phi l/M_0}}{M_0}  \frac{\sin \pi l f_0}{\sin( \pi l /M_0)}.
\end{equation}
Using that $K = \sqrt{v_J/v_N}$ and $v_s = \sqrt{v_N v_J}$~\cite{C04b,Gia04}
and,  to leading order in $\gamma^{-1}_L$, $E_0 = \langle G | H_0 | G \rangle + \langle G| H_{\rm int}| G\rangle$,
where $|G\rangle$ is the ground state of $H_0$, along with (\ref{stiff}), in the $M_0\to\infty$ 
Eqs. (\ref{K}) and (\ref{vs}) are obtained (the same expressions were also derived by carefully taking the field-theoretic 
continuum limit of $H_{\rm F}$~\cite{unpub}).

   Before concluding,  an important difference between the Bose-Hubbard 
and Lieb-Liniger models is worth discussing: whereas the latter displays Galilean invariance, which
implies that $v_J$ must be equal to the Fermi velocity $v_F$~\cite{C04b}, in the former
this symmetry 
is broken by the lattice.  By inspection of Eq.~(\ref{hf}) it can be seen 
that the terms responsible for the violation,
that is, for the renormalization of $v_J$ away from $v_F$ are those
those coming from $H_1$. 
Thus, the renormalization of $v_J$ becomes manifest after 
noticing  that $[H_1, n_m] \neq 0$, and therefore one expects
a non-zero contribution to the coefficient of $\partial_x j(x,t)$
($j(x,t)$ being the long wave-length part of the
current density) in the coarse-grained continuity equation~\cite{Gia04}. 
This is one  notable feature of  $H_{\rm F}$, which is obtained by  projecting
on a low-energy subspace where $n_m \leq 1$~\cite{C03}, in what may be regarded
as a first step of the renormalization group.

 To sum up, we have shown that the Tonks regimes in the continuum and
 on the lattice are not, strictly speaking, equivalent. The local two-body correlations
 of the system on a lattice saturate with increasing temperature while for the Tonks
 regime of the continuum model are known to increase monotonically~\cite{K02,C03}. Furthermore,
 the parameter $K$, characterizing the decay of long-distance correlations~\cite{C04b,Gia04}, 
 is more easily tuned to the Tonks limit on the lattice than in the continuum. Finally, our results
can also be extended to  the calculation of corrections to the internal 
energy and entropy of the lattice gas~\cite{unpub}.
 However, the distinct behavior of the two Tonks limits is  well displayed by  the properties
 considered in this work, and we shall not pursue this task here. 
 This research has been supported  through a \emph{Gipuzkoa} fellowship
 granted by \emph{Gipuzkoako Foru Aldundia} (Basque Country).

\vskip 5mm

$\bullet$ {\bf Appendix: On the equivalence of the first and second quantization 
approaches to fermionization}. The connection between fermionization in the wave function
formalism and its second quantization version, which leads to effective Hamiltonians like
Eq.~(\ref{hferm}), has not been sufficiently emphasized in previous treatments (\emph{e.g.}
Ref.~\onlinecite{C03}). For completeness, we  include a proof of their equivalence
in this appendix.

  In their pioneering 1928 paper on second quantization of fermion fields, Jordan and Wigner~\cite{JW28}  introduced a transformation from hard-core bosons (or Pauli matrices) to fermions:
\begin{eqnarray}
b_m &=& K_m c_{m} \quad b^{\dag}_{m} = c^{\dag}_m K_m, \label{jw1}\\
K_{m} &=& \exp\left[ i\pi \sum_{l<m} n_l \right] = \prod_{l<m}\left( 1 - 2n_l \right). 
\end{eqnarray}
The operator $K_m$ (often referred to as the Jordan-Wigner string)  turns the  
(hard-core) boson operator $b_m$ into the fermionic  $c_m$  by attaching to it a
phase factor which is determined by the number of particles to the left of  site $m$.
The trick converts the commutation relations of the $b$'s at different sites  into the anti-commutation
relations of the $c$'s. In this appendix, it is shown that the same trick yields
the celebrated Bose-Fermi mapping due to Girardeau~\cite{G60}. Let us consider the 
N-particle bra:

\begin{equation}
|\Phi \rangle = \sum_{\{m_i = 1 \}_{i=1}^N}^{\{M_0\}} \Phi_F(x_{m_1},  \ldots, x_{m_N} )\: c^{\dag}_{m_1}
\cdots c^{\dag}_{m_N} | 0 \rangle, \label{wf}
\end{equation}
where $| 0 \rangle$ is the empty state. The wave function $\Phi_F(x_1,\ldots,x_N)$ is 
anti-symmetric under exchange of any pair of coordinates
as a result of the anti-commutation of
the $c$'s. This may lead us to think that the above bra
describes a system of $N$ fermions.  However, by noticing 
that $K^{-1}_m = K_{m}$, we can invert 
(\ref{jw1}) and write the product
\begin{equation}
 c^{\dag}_{m_1}
\cdots c^{\dag}_{m_N} | 0 \rangle =  b^{\dag}_{m_1} K_{m_1} \cdots b^{\dag}_{m_N} K_{m_N} |0 \rangle.
\end{equation}
Next we shift all the string operators to the right and use that $K_{m} | 0\rangle = |0\rangle$ (since
$n_l |0 \rangle = 0$ for all $l$) every time a string operator hits the empty state. Nevertheless, when
commuting a string operator with a creation operator one must take care of a phase factor:
$K_n  b^{\dag}_m =  e^{i \pi \theta(x_n -x_m )}\:  b^{\dag}_{m} K_n$ (where $\theta(0) = 0$ is assumed).
After shifting all the string operators to the right, a factor like this one appears  
for each pair of particles, and therefore,
\begin{equation}
 c^{\dag}_{m_1}
\cdots c^{\dag}_{m_N} | 0 \rangle = A(x_{m_1},\ldots, x_{m_N}) b^{\dag}_{m_1} \cdots b^{\dag}_{m_N} | 0 \rangle,
\end{equation}
where the  fully antisymmetric prefactor $A(x_{m_1},\ldots,x_{m_N}) =  e^{i\pi \sum_{i<j} \theta(x_{m_i}-x_{m_j})}= \prod_{i<j} {\rm sgn}(x_{m_i}-x_{m_j})$. Introducing the last expression into (\ref{wf}), the bra can be rewritten as
\begin{equation}
| \Phi \rangle  =  \sum_{\{m_i = 1\}_{i=1}^N}^{\{M_0\}} \Phi_B(x_{m_1},  \ldots, x_{m_N} )\: b^{\dag}_{m_1}
\cdots b^{\dag}_{m_N} | 0 \rangle,
\end{equation}
where
\begin{eqnarray}
\Phi_B(x_{m_1}, \ldots, x_{m_N})&=& A(x_{m_1}, \ldots, x_{m_N}) \Phi_F(x_{m_1},\ldots,x_{m_N}) \nonumber \\
&=&   |\Phi_F(x_{m_1},\ldots,x_{m_N})|
\end{eqnarray}
is  a symmetric function which vanishes if $x_{m_i} = x_{m_j}$ for $i\neq j$. In other words,
$\Phi_B$ is the wave function of a system of hard-core bosons. This proves the equivalence of the
first and second quantization approaches to fermionization.

\end{document}